**Societal and scientific impact of policy research: A large-scale empirical study of some explanatory factors using Altmetric and Overton**

Pablo Dorta-González [1,*], Alejandro Rodríguez-Caro [2], María Isabel Dorta-González [3]

1 Institute of Tourism and Sustainable Economic Development (TIDES), Campus de Tafira, University of Las Palmas de Gran Canaria, 35017 Las Palmas de Gran Canaria, Spain ORCID: http://orcid.org/0000-0003-0494-2903

2 Department of Quantitative Methods in Economics and Management, Campus de Tafira, University of Las Palmas de Gran Canaria, 35017 Las Palmas de Gran Canaria, Spain; alejandro.rodriguez@ulpgc.es; ORCID: 0000-0002-8080-3094

3 Department of Computer and Systems Engineering, Avenida Astrofísico Francisco Sánchez s/n, University of La Laguna, 38271 La Laguna, Spain; isadorta@ull.es; ORCID: 0000-0002-7217-9121

* Correspondence: pablo.dorta@ulpgc.es

**Highlights:**

- The study provides insights into the dynamics of policy impact.
- We use principal component analysis to show the unique patterns of different stakeholders.
- OLS regression modelling is used to quantify the effect size of influences on policy citations.
- The findings have implications for policy actors and strategies to increase policy impact.

**Abstract**

This study investigates how scientific research influences policymaking by analyzing citations of research articles in policy documents (policy impact) for nearly 125,000 articles across 434 public policy journals. We reveal distinct citation patterns between policymakers and other stakeholders like researchers, journalists, and the public. News and blog mentions, social media engagement, and open access publications (excluding fully open access) significantly increase the likelihood of a research article being cited in policy documents. Conversely, articles locked behind paywalls and those published under the full open access model (based on Altmetric data) have a lower chance of being policy-cited. Publication year and policy type show no significant influence. Our findings emphasize the crucial role of science communication channels like news media and social media in bridging the gap between research and policy. Interestingly, academic citations hold a weaker influence on policy citations compared to news mentions, suggesting a potential disconnect between how researchers reference research and how policymakers utilize it. This highlights the need for improved communication strategies to ensure research informs policy decisions more effectively. This study provides valuable insights for researchers, policymakers, and science communicators. Researchers can tailor their dissemination efforts to reach policymakers through media channels. Policymakers can leverage these findings to identify research with higher policy relevance. Science communicators can play a critical role in translating research for policymakers and fostering dialogue between the scientific and policymaking communities.

## 1. Introduction

Scientific research has value not only in terms of its academic impact, typically measured by citation metrics, but also in terms of its broader societal impact. Societal impact, or impact on society, includes the impact of research on all sectors of society (possibly excluding the impact of research on science). In a study by Wilsdon et al. (2015), the societal impact of research was defined as its impact on education, society, culture, or the economy. One approach is altmetrics, which provides a quantitative means of measuring the broader impact of publications, as highlighted by the NISO Alternative Assessment Metrics (NISO, 2016).

As digital scholarly communication has evolved, there has been a significant shift in how the societal impact of scholarly research is evaluated. This has resulted in a more diverse approach, encompassing a wider range of scholarly publications and creative communication methods (see surveys by Bornmann, 2013; de Rijcke et al., 2016; Bornmann and Haunschild, 2019). To assess the quality of research being conducted at higher education institutions, the UK implemented the Research Excellence Framework (REF). Within this framework, the evaluation of impact beyond the scientific realm holds significant weight, accounting for 25% of the total assessment. This includes measuring the influence of research on public policy, services, economy, society, culture, health, environment, and overall quality of life (see Khazragui and Hudson, 2015).

In the early years of altmetrics research (the first half of the 2010s), many studies insisted on counting the online visibility of papers and linking it to the citations of papers, as if measuring impact using citations and altmetrics mentions were somehow equivalent. However, all of these studies lack a theoretical rationale for why we should expect such a connection and positive relationship between societal relevance and scientific impact (see Ravenscroft et al., 2017). The literature on research evaluation methods has long emphasized that societal impact and relevance beyond academic boundaries arise differently from scientific impact (see Spaapen and van Drooge, 2011; Joly et al., 2015; Morton, 2015).

The initial enthusiasm for altmetrics as a proxy for societal impact was probably justified, given the increasing emphasis on societal impact by funding agencies. However, there is a growing recognition in the altmetrics literature that it is time to rethink how these measures are used and understood. In recent years, several studies have challenged this approach and suggested that altmetrics should be reframed as indicators of science-society interactions and the dissemination of knowledge beyond academic boundaries, rather than direct measures (counts) of impact (see Haustein et al., 2016; Robinson-Garcia et al., 2018; Díaz-Faes et al., 2019; Costas et al., 2021; Wouters et al., 2019; Alperin et al., 2023).

The integration of scientific articles into policy documents is a powerful indicator of the impact of research on society (Yu et al., 2023). In addition, the citation of research in policy documents increases the credibility of the authors referenced and of the policy documents themselves. As revealed by Bornmann et al. (2016), this approach provides valuable insights into the interconnectedness of academic research and policymaking.

Despite the longstanding ideal of evidence-based policymaking, a gap persists between the production of scientific knowledge and its utilization in policy development. This study addresses this critical juncture by investigating the factors influencing how policymakers reference research. By uncovering the distinct citation practices of policymakers and other stakeholders, this research sheds light on the communication channels most impactful for bridging the knowledge gap between academia and policy. Furthermore, the finding that academic citations hold less sway over policymakers than news mentions compels a critical reevaluation of knowledge dissemination strategies. This study offers valuable insights for researchers, policymakers, and science communicators seeking to optimize the translation of scientific knowledge into actionable policy decisions.

## 2. Literature review

*Linking research and policy*

According to the findings of a study conducted by Willis et al. (2017), citation was identified as one of the top eight influential factors examined by stakeholders when determining social outcomes in policy papers. Another study by Yin et al. (2021) analyzed how science and policy intersected in the context of the COVID-19 pandemic using data from Overton. Their research revealed that a significant number of policy papers related to the pandemic heavily relied on current and influential peer-reviewed scientific studies. Additionally, in the policy area, publications that referenced scientific research were highly valued. Overall, national governments tended to indirectly acquire and utilize scientific knowledge through intergovernmental organizations (IGOs), although the extent to which science was used in policymaking varied among different institutions. Notably, the World Health Organization (WHO) played a significant role in mediating the transfer of scientific knowledge to policy decisions. Also, in the context of COVID-19, Dorta-González (2022) examined the relationship between citations in policy documents and mentions on Twitter, finding a significant positive correlation between both variables according to Spearman's test.

In a recent study, Bornmann et al. (2022) delved into the intricate ties between policy and research surrounding climate change. Their findings revealed that articles referenced in policy documents pertaining to climate change garnered significantly higher citation counts than those not referenced. Furthermore, their model shed light on the diverse ways in which scientific research impacts policy development across various types of papers.

Conversely, in their quest to determine the practicality and influence of research, Newson et al. (2018) implemented a retroactive tracing method to assess 86 policy documents from New South Wales focusing on childhood obesity. Their conclusions indicated that, in this instance, the policy papers' mentions of the cited research were insufficient to prove its impact on the policy-making process.

In addition, a meta-research study by Abbott et al. (2022) of early COVID-19 published evidence syntheses sought to explore the relationship between the quality of reviews and the level of interest from researchers, policymakers and the media. Although only a limited number of reviews were cited in policy documents, the study raised concerns that the quality of reviews did not significantly influence their citation in policy documents.

*Coverage of policy document data aggregators*

The Altmetric Attention Score, developed by Altmetric.com, serves as an indicator of the total attention a research result has received. Each source contributing to this score is given a weight, with policy documents having a weight of 3 (compared to 8 for news, 5 for blogs, 3 for Wikipedia and 0.25 for Facebook and X - formerly Twitter, posts and re-posts).

Previous research has shown that citations of research in policy documents do not receive substantial coverage in the Altmetric.com dataset, compared to mentions on Twitter and Facebook. Bornmann et al. (2016) found that only 1.2% of research papers received at least one citation in policy documents. Haunschild and Bornmann (2017) reported that less than 0.5% of papers in different subject categories were cited at least once in policy-related documents.

According to a study conducted by Tattersall and Carroll (2018) at the University of Sheffield, a mere 1.41% of the 96,550 research outputs tracked by Altmetric.com were cited in policy documents. This low rate raises concerns about the effectiveness of multidisciplinary research in influencing policy decisions. In light of this, a recent study by Szomszor and Adie (2022) has shown that the new altmetrics database, Overton, provides more comprehensive coverage of policy document citations compared to Altmetric.com. This highlights the importance of using accurate and thorough data sources when evaluating the impact of research on policy. In an effort to assess the connection between multidisciplinary research and its adoption in policy papers, Pinheiro et al. (2021) matched the Overton database with Scopus data. Their findings revealed a coverage rate of 6.0% for all funded publications in the dataset, indicating that a greater percentage of multidisciplinary research.

Haunschild and Bornmann (2017) highlight several factors that contribute to the lack of research citations in policy documents: (1) One of these factors is the limited scope of Altmetric.com's policy document sources, which has not yet garnered enough data for comprehensive coverage. (2) Because most of the literature is written primarily for academic audiences, only a small portion of it may be actually relevant to policy. (3) Policy documents might not use a scientific citation style and their authors, who are frequently not researchers, may only occasionally include scientific studies in their writing. (4) There can be obstacles and little communication between policymakers and researchers.

*Motivations for engagement with other altmetrics*

The motivation behind mentions of research in policy documents is an area of research that has not received much attention. However, there have been numerous studies that have focused on understanding the motivations behind engagement on various social platforms such as blogs, Facebook, Twitter and Sina Weibo. These studies have predominantly used interview and content analysis methods to explore these motivations.

Shema et al (2015) used content analysis to classify blog post material in the health category of Researchblogging.org, categorizing motivations into broad themes of discussion, critique, guidance, trigger, extension, self, controversy, data, ethics, and other. Notably, the most commonly cited motivations were guidance, critique and conversation.

Academics have also explored the reasons for disseminating scholarly work on Twitter. According to Veletsianos (2012) analysis of tweets from 45 academics, sharing knowledge, resources and media was the most common purpose. Na (2015) conducted a content analysis to investigate the reasons for English tweets referencing academic publications in the field of psychology. Discussion was found to be the main motivator, and a significant subset of motivation was devoted to describing and interpreting scientific findings. Yu et al (2017) used content analysis in the Sina Weibo environment to identify four main drivers of scientific engagement: discussion, marketing, triggering and distribution.

Furthermore, an online poll was carried out by Syn and Oh (2015) to look at the elements that motivate people to share information on Facebook and Twitter. Motivators like enjoyment, efficacy, learning, self-gain, compassion, empathy, social engagement, community interest, reputation, and reciprocity were identified by their research. On these platforms, users share information with the expectation of feeling involved and connected to online communities.

The literature review reveals motivations for sharing scholarly work online: sharing knowledge, engaging in discussion, seeking guidance, and fostering connections. Purposes such as guidance, critique and conversation are common, reflecting a desire for dialogue and feedback exchange. Motivations include enjoyment, efficacy, learning and social engagement, suggesting that online sharing fulfils informational and social needs. However, it's important to recognize the limitations of the methodologies and sample populations of these studies. While they provide valuable insights, the findings may not be fully representative of broader online communities. In addition, biases in content analysis and online surveys may affect the interpretation of the results. Nevertheless, these studies offer valuable implications for understanding the motivations behind online information sharing, which can inform strategies for effective communication and knowledge dissemination in online environments. By considering these findings, researchers can tailor their approaches to better engage with their audiences and maximize the impact of their work.

## 3. Methodology

*The aim of the study (objective)*

Our study aims to fill a gap in the existing literature by investigating the factors driving research citations in policy documents, an area that has received limited attention. While previous research has explored the motivations behind research citations on social media platforms, there remains a lack of understanding of the determinants of citations in public policy documents. To address this gap, we used Principal Component Analysis (PCA) to identify related variables and conducted a regression analysis (with robust ordinary least squares estimators) to explore the potential effects of different types of influence and bibliometric variables on citations in public policy. This comprehensive model also includes other relevant factors such as access type, journal impact factor and funding sources. Through this multifaceted approach, our study aims to provide valuable insights into the complex dynamics underlying the dissemination and use of research in public policy.

*Data*

We used the Web of Science core collection database to identify journals indexed as of July 2023 that contain the term "policy" in their titles. This resulted in the identification of 434 journals specializing in public policy. Through the ISSN of said journals, a total of 124,778 research articles with DOIs were located in the Altmetric and Overton databases. The entire data collection took place in July 2023.

Regarding the existence of various methods for measuring non-academic research impact, we opted to focus on citations within policy documents for several reasons. Firstly, this approach directly targets the policy sphere, enabling us to identify research articles that demonstrably influence policy development. Therefore, our method provides a clearer picture of the real-world application of scientific evidence. Secondly, policy citations offer a quantifiable measure of impact, allowing for statistical analysis and comparisons across different research fields and policy areas. This facilitates the identification of trends and patterns in how policymakers utilize research.

However, we acknowledge limitations inherent to this approach. Policy documents might not capture all instances where research informs policy. Additionally, our reliance on citations within these documents introduces potential biases. For example, some policy areas might be more likely to cite research than others. To address these limitations, future studies could employ complementary methods. Interviews with policymakers or analyses of policy deliberations could provide a more nuanced understanding of research-policy interactions.

Concerns about the representativeness of the analyzed sample arise from the multidisciplinary or interdisciplinary nature inherent in the field of public policy. This trait suggests that many policy studies may be published in journals that do not explicitly feature the term 'policy' in their titles. Therefore, the lack of such studies is likely to be unevenly distributed across various policy domains. Nevertheless, as we show below, there are compelling arguments to support the idea that such a sample can provide valuable insights into public policy.

Journals that explicitly include 'policy' in their titles usually focus primarily on policy-related themes. This specialization means that the articles published in these journals are more likely to be directly relevant to public policy, which enhances the representativeness of the sample in terms of topic specificity. Additionally, these journals follow editorial boards and peer review processes that are committed to ensuring the quality and relevance of published content in the field of public policy. These mechanisms help to filter out irrelevant or substandard articles, thus improving the representativeness of the sample in terms of content reliability and rigor.

Furthermore, publications that focus specifically on policy topics often have greater visibility and prestige within the public policy community. Authors and researchers may prioritize these outlets for disseminating their work, resulting in a higher concentration of influential and impactful articles within the sample. This aspect enhances the representativeness of the sample by including significant contributions to the field. Policy-focused journals also serve as hubs for networking and community engagement among policymakers, researchers, and practitioners. Therefore, they are more likely to attract a diverse range of authors and perspectives, contributing to the representativeness of the sample in terms of demographic and ideological diversity.

Finally, reputable journals that have a history of publishing policy-related content may have gained trust and credibility within the field over time. Researchers may view these publications as reliable sources of information and give them priority when submitting their work, further enhancing the representativeness of the sample in terms of historical continuity and tradition.

For each of the 434 public policy journals, we collected basic information including ISSN, e-ISSN, Journal Impact Factor (JIF), and indexing categories in the Web of Science database. We also collected at the paper level the access type (gold, hybrid, green, bronze, closed), publication date, and funding (funded and unfunded).

The categorization of access types was extracted directly from the paper itself, rather than being inferred from the journal. Both the Altmetric and Overton databases capture access types at the document level, allowing for nuanced distinctions among papers within the same journal, volume, and issue. For instance, a journal issue may include papers with various access typologies, such as paywall access (OA closed), open access provided by the journal's publisher to increase citations and journal impact (OA bronze), or open access facilitated through thematic or institutional repositories (OA green).

It is important to acknowledge that the availability of various access types reflects the changing practices of scholarly communication and initiatives to enhance the accessibility and dissemination of research outputs. Each access type has different implications for readership, visibility, and scholarly impact. Paywall access typically limits access to paying subscribers, while various forms of open access aim to increase accessibility by removing financial or technical barriers. The diversity of access types within a single journal highlights the complexity of contemporary scholarly publishing and the significance of considering access dynamics in bibliometric analyses and research evaluation frameworks.

We developed the typology for categorizing policy journals through a systematic process involving the examination of several sources of information. First, we identified relevant indexing categories and examined the titles and descriptions of policy journals available in academic databases. We then reviewed information available on journal websites to further refine the classification criteria. Through this iterative process, we produced a comprehensive set of 12 pre-defined policy typologies covering a wide range of policy areas. To ensure the reliability of the categorization process, two authors independently undertook the task of classifying journals into the predefined typologies.

Each of the 12 policy typologies was defined on the basis of different characteristics and areas of focus within policy research. For example, agricultural policy journals focus primarily on issues related to agricultural practices, policies and their impacts. Cultural policy journals focus on issues related to cultural heritage, arts and cultural management. Similarly, economic policy journals deal with issues of economic development, trade and fiscal policy, while education policy journals focus on education reform, pedagogy and equity. Each typology has been carefully delineated to capture the specific area of policy research it represents.

The distribution of the 12 policy typologies across the journals was as follows: agriculture policy (4), cultural policy (6), economic policy (92), education policy (34), environmental policy (60), foreign policy (28), general (16), healthcare policy (57), legal policy (28), public administration policy (24), social policy (71), and technology policy (14).

Our research employed a comprehensive strategy to identify references to policy research across a variety of online platforms, including Wikipedia and social media networks. Using the Altmetric.com database, we accessed indexed articles via their journal International Standard Serial Number (ISSN) and electronic International Standard Serial Number (e-ISSN), which allowed us to locate articles with Digital Object Identifiers (DOIs) in Crossref, along with associated altmetrics. In cases where DOI information was not available, we chose to exclude the paper from our dataset to ensure the integrity and accuracy of our analysis.

This thorough methodology resulted in a substantial dataset of 124,778 articles with DOIs, of which 100,703 received some form of social attention (excluding citations from the Dimensions database and readership statistics from the Mendeley reference manager.) Notably, the vast majority of these articles - around 80,000 - were published within the last decade, which represents the beginning of data collection from multiple sources. Finally, we also used the DOIs to locate these articles in the Overton.io database and extracted the number of citations in policy documents.

*Policy impact indicators*

This study delves into the complexities of measuring policy impact by strategically utilizing two complementary data sources: Overton and Altmetric. While both indicators aim to capture how research influences policy, they offer distinct perspectives. Overton serves as a vast and comprehensive index, encompassing policy documents, guidelines, think tank publications, and working papers from over 182 countries and exceeding 1,000 global sources. This expansive reach offers several advantages. Overton's high volume allows for the identification of a broader range of policy citations. This is particularly valuable for research that might have a more localized or niche policy influence, potentially missed by other sources. Furthermore, by incorporating data from a multitude of countries, Overton provides insights into how research informs policymaking across diverse geographical contexts. This broadens our understanding of the international reach of research impact.

However, Overton's focus on high volume also presents a potential limitation. The extensive number of documents might encompass some that reference research in a less substantial way, potentially diluting the signal of direct policy influence. Therefore, this study leverages the strengths of both Overton and Altmetric. The Principal Component Analysis (PCA) with both data sources combined helps assess the overall coherence of the findings, while the separate Ordinary Least Squares (OLS) regression analysis allows for a deeper dive into the unique contributions of each indicator. This multifaceted approach strengthens the validity of the results and provides a more complete picture of how research shapes policy decisions.

*Methods*

In terms of our statistical methodology and its application, we chose to use Spearman correlation rather than Pearson correlation, mainly due to the characteristics of our data. Spearman correlation is a non-parametric measure of association that assesses the strength and direction of monotonic relationships between two variables. Unlike Pearson correlation, Spearman correlation does not assume a normal distribution or linear relationships, making it suitable for variables that may not fulfil these assumptions. As a result, Spearman's correlation provides a robust measure of association that does not depend on these assumptions and is better suited to the aims of our research.

In the regression analysis, we used a form of linear regression known as ordinary least squares (OLS). OLS regression is a widely used technique for exploring the relationship between a dependent variable and one or more independent variables by estimating coefficients that best fit the observed data points. Regression analysis allowed us to assess the extent to which different predictors were associated with the outcome variable (number of citations in policy documents), while controlling for other relevant factors. OLS regression is particularly useful when a linear and additive relationship between variables is assumed, which seems reasonable in the context of our study. Overall, regression analysis facilitated our ability to quantify the influence of different factors on citations in policy documents while controlling for potential confounding variables.

As indicated, we used robust ordinary least squares estimators, incorporating White's robust parameter variance estimation technique. This is a statistical technique commonly used in econometrics to compute variance estimates of regression model coefficients. This approach is particularly valuable when faced with problems such as heteroskedasticity (non-constant error variability) and autocorrelation in regression data. By adjusting the variance estimates, White's robust parameter variance estimation helps to correct for potential problems that might otherwise compromise the precision of standard estimates, ensuring more reliable statistical inference in the presence of departures from the classical assumptions of linear regression.

To ensure balanced contributions from all attributes and to avoid bias towards any one variable, we normalized the data by subtracting the mean and dividing by the standard deviation for each variable. In a geometric context, eigenvectors represent directions while eigenvalues quantify the variance along those directions. Each eigenvector is paired with an eigenvalue, with the highest eigenvalue corresponding to the first principal component, followed by the second principal component corresponding to the eigenvector with the second highest eigenvalue, and so on. Transforming the data into a new dimensional space involves repositioning the data within a subspace defined by these principal components. This reorientation is achieved by multiplying the original data by the previously computed eigenvectors. Importantly, this transformation preserves the original data while providing a new perspective that enhances its representation.

To measure the overall impact of research, we can consider three aspects. The first is academic impact, which is based on how often other researchers cite the research and how much it influences their work. The second is societal impact, which concerns how the research affects the wider world, such as its use in policymaking and innovation, as demonstrated by citations in policy documents and patents. The third is social attention and its impact on public discourse, which can be measured through altmetrics. Altmetrics include news mentions, blog discussions, social media engagement, and references on platforms like Wikipedia. These metrics demonstrate how research reaches and engages with diverse online audiences. For instance, if a research study receives significant attention on social media or is extensively covered by mainstream news outlets, it indicates that the research has piqued public interest and is impacting broader societal discussions. Monitoring these indicators enables researchers and institutions to comprehend the significance of their work beyond conventional academic metrics.

Social attention can be categorized into three types: media attention, social media attention, and encyclopedic attention. Encyclopedic attention aims to disseminate knowledge through sources such as online encyclopedias or specialized websites. This type of attention differs from the other types because it focuses on providing accurate and comprehensive information on different topics, with an educational and referential purpose. Media attention refers to media coverage of relevant events and news, while social media attention involves interaction and participation on digital platforms. Encyclopedic attention, on the other hand, aims to inform and educate the public through well-organized and well-researched content, as seen in encyclopedic definitions and developments in dictionaries and encyclopedias.

Therefore, in this work we examined the following categories of influences: (1) academic impact, assessed via scholarly citations in the Dimensions database; (2) policy impact, evaluated through policy citations in the Altmetric and Overton databases; (3) media attention, measured by online mentions in news and blogs; (4) social media attention, determined by mentions on Twitter and Facebook; and (5) encyclopedic attention, identified through mentions in Wikipedia.

Our approach aggregates news articles and blog posts. Traditional press tends to focus on the most sensational aspects of science, which appeal to a large portion of society. As a result, some areas of research receive minimal media coverage. To address this limitation, scientists often act as science communicators through specialized blogs. We decided to combine both communication channels, despite their differences, because they both aim to reach a wider audience that may not be familiar with the subject matter.

## 4. Results

*Statistics and Spearman correlations*

The distributions of the variables are skewed and loaded with zeros (see Table 1). The presence of both characteristics, skewness and a significant number of zeros, complicated statistical analyses and required specialized methods, which are described in the methodological section. Figure 1 serves as a comprehensive visual representation, illustrating that the extent of coverage within the two policy databases, namely Altmetric on the left and Overton on the right, remains disparate regardless of the type of public policy under consideration. In particular, Overton demonstrates a superior ability to aggregate mentions within policy documents compared to Altmetric.

In addition to highlighting the disparity in performance between Altmetric and Overton, Figure 1 sheds light on a crucial feature inherent to this dataset. Looking closely at the figure, in the early stages, corresponding to lower values along the horizontal axis, there is a proliferation of data points accompanied by a significant degree of variability. Conversely, as one moves towards higher values on the same axis, a contrasting pattern emerges: the number of data points decreases, and the variability decreases significantly. This observation implies that in the field of public policy and its associated datasets, there is a dynamic relationship between the variable of interest and the abundance and variability of data points. Such insights are crucial for a nuanced understanding of the patterns and trends within policy documents, contributing to a more informed and comprehensive analysis of the data at hand.

As shown in Figure 2, it is clear that, except for policy citations in Altmetric and Overton, the Spearman correlations are generally of low magnitude. It is worth noting the discernible negative correlation between social media mentions and academic citations, although this correlation is relatively modest. Looking at the disaggregated information by policy type shown in Figure 3, different patterns emerge, but overarching trends persist. In particular, there is a notable difference when looking at Wikipedia mentions, which show a discernible negative correlation with social media mentions in the context of foreign policy, legal policy, administrative policy and education policy. There is also a small correlation between academic citations in cultural policy and Wikipedia mentions.

*Consolidation of correlated variables: Principal Component Analysis (PCA)*

By combining correlated variables and reducing the dimensionality of our high-dimensional data, we were able to uncover key insights. Table 2 presents the six principal components that correspond to the number of variables in the dataset, denoted as PCA1 through PCA6. Each of these elements contributes to the explanation of a fraction of the dataset's overall variation. The first principal component in the cumulative proportion section explains 28% of the total variance. This indicates that the first principal component alone can effectively capture more than a quarter of the data across all six variables. Roughly 21% of the total variation is explained by the second main component. The first two principal components can account for half of the dataset, according to the cumulative proportion of PCA1 and PCA2. The third principal component explains 17% of the total variation in the dataset. Looking at the cumulative proportion of the first three principal components (PCA1, PCA2 and PCA3), together they capture over 67% of the variation. This means that the first three components alone can capture more than two-thirds of the variability in the data. This finding is significant as it suggests that even with only three components, a significant amount of information can be retained after reducing the dimensionality of the dataset.

The optimal number of components for a PCA may vary depending on the dataset and the objectives of the analysis. It is imperative to find the balance between avoiding excess, which may introduce noise or redundant information, and including the necessary number of principal components to capture the desired variance. One approach to gaining insight into their importance is to examine their loadings in the principal components. These loadings indicate the relationship between the original variables and the principal components. This loading matrix highlights the coefficients of the original variables that make up each principal component, providing a clearer understanding of their relationship.

The loading matrix of the first three principal components (PCA1, PCA2 and PCA3) for different factors related to mentions and citations is shown in Table 3. A variable is represented by a row and its loading on a corresponding principal component is represented by a column. For example, looking at PCA1, we see that mentions in blogs and news have a loading of -0.32, mentions in social media have a loading of -0.23, and so on. These loadings indicate the direction and strength of the relationship between each variable and its corresponding principal component. For each variable, PCA2 and PCA3 also have their corresponding loadings.

The loadings matrix in Table 3 draws attention to the negative values in the first main component, particularly the strong negative values associated with policy citations. It can be concluded that papers with higher citations and mentions tend to have lower PCA1 scores. As a result, PCA1 could be classified as a basic/applied research dimension. Note that basic research is not motivated by the need to produce useful solutions or applications to real-world problems, but is usually driven by curiosity and the desire to understand underlying principles and phenomena, with no immediate or direct application in mind.

Looking at the loading matrix with a focus on the second principal component, policy citations in Altmetric and Overton are correlated with positive values. This suggests that articles with more policy citations are more closely associated with this second principal component. PCA2 can therefore be classified as a policy dimension. In addition, the loading matrix for the third principal component shows positive values associated with mentions in social media, news and blogs. This suggests that there is a stronger correlation between this third principal component and works that have a higher frequency of social mentions. PCA3 can therefore be recognised as a sociological dimension.

Figures 4 and 5 show the contribution of the variables to the first principal components. The use of these plots facilitates a visual understanding of the relationships and distinctions between the variables and provides insights into the influence of each attribute on the respective principal components. Figure 4 provides two key insights. First, it shows that variables that are grouped together have positive correlations with each other, as is the case with policy citations in Altmetric and Overton, or news and blog mentions with social media mentions. Similarly, there is a positive correlation between Wikipedia mentions and academic citations. Second, the distance between a variable and the origin of the coordinates in the biplot reflects its level of representation. In particular, policy citations have the largest magnitude in Altmetric and Overton, indicating strong representation. In addition, news and blog mentions, social media mentions and Wikipedia mentions have larger magnitudes than academic citations, suggesting that they are better represented than academic citations.

Figure 4 also attempts to assess the degree to which each variable contributes to a particular component, as measured by the square cosine (cos2). A low cos2 value indicates poor representation, while a high value indicates strong representation. In Figure 4, variables with similar cos2 values are colour coded accordingly. Variables with high cos2 values are shown in green, including policy citations in Altmetric and Overton. Medium cos2 variables are shown in orange, such as news and blog mentions. Finally, variables with low cos2 values are shaded black, including social media mentions, Wikipedia mentions and academic citations.

Similar findings on the association between variables in a three-dimensional space can be visualised in Figure 5. This plot shows the persistence of relationships between variables that extend into the third dimension. As a result, variables that cluster together show positive correlations, exemplified by cases such as political citations in Altmetric and Overton. Similarly, there is a positive correlation between Wikipedia mentions and academic citations, and between news and blog mentions and social media mentions.

This may have the following implications for the dynamics of bibliographic referencing by different actors, at least in the specific context of public policy. It is important to note that the data analysed refer to citations and social mentions of articles published in public policy journals indexed in the Web of Science database:

(1) The dynamics of bibliographic referencing by the authors of policy documents (policymakers) are significantly different from those of other stakeholders.

(2) The dynamics of bibliographic referencing by researchers writing scientific papers and other authors contributing to encyclopaedic articles on Wikipedia are relatively similar to each other, but significantly different from the other stakeholders.

(3) The dynamics of bibliographic referencing by individuals writing news articles in journalistic media and posts on blogs communicating scientific results (journalists and science communicators) and by the general population through social networks are relatively similar to each other, but significantly different from the rest of the stakeholders.

*Effects of different influence types and bibliometric variables on policy citations: Ordinary Least Squares Regression*

We consider two explanatory models, one for each source of citations in public policy documents (Model 1: Y = policy citations in Altmetric; Model 2: Y = policy citations in Overton). The explanatory variables (Xi) include the following: news and blog mentions, social media mentions, Wikipedia mentions, academic citations, access type (gold, hybrid, green, bronze, closed), journal impact factor, funding (funded and unfunded), year of publication, and type of policy.

The rationale behind the selection of these explanatory variables lies in the overall objective of the study, which is twofold. Firstly, it seeks to explore the intricate interrelationships between the different types of impact and thereby shed light on their complex dynamics. Secondly, it seeks to provide a thorough analysis of the different knowledge referencing practices of different stakeholders involved in the dissemination of knowledge in public policy. By carefully examining these variables, the study aims to provide nuanced insights into the multifaceted landscape of policy communication and its impact on decision-making processes.

The model hypothesis regarding causality is closely linked to the temporal sequence of key events in the dissemination and reception of research feedback. The timing of these events not only reflects the dynamic trajectory of scientific impact, but also serves as a valuable indicator of the evolving influence of a scientific paper. As soon as a new paper is published, a flurry of activity begins as news items and online postings are produced and distributed. This initial phase signals the research's entry into public discourse and attracts the attention of different audiences. These early stages lay the groundwork for subsequent waves of engagement and the wider impact of the paper.

The second major milestone in this timeline is the proliferation of mentions on social media platforms. During this phase, the research is woven into the digital landscape of these networks, with a particular focus on the first few weeks after publication. This not only extends the reach of the paper, but also sparks dynamic discussions among readers around the world, facilitating early engagement and debate around the concepts and findings presented. The third stage of the timeline is marked by the emergence of scholarly impact through citations in scholarly journals. This typically occurs around six months after the initial publication and signifies a deeper integration of the study into the academic discourse. The number of citations a publication receives serves as a tangible measure of its influence on subsequent research, highlighting its role in shaping and advancing the field.

When policy documents and Wikipedia articles incorporate the bibliographic references, this chronological trend reaches its conclusion. This final stage represents a strong endorsement and validation of the study, both within and beyond the scientific community. The paper's findings are incorporated into reputable knowledge databases and policy deliberations, underlining its enduring importance and impact on the wider understanding of the issue.

This model proposes a comprehensive framework for assessing the impact of a scientific paper on both public discourse and academic scholarship. This framework involves analyzing the temporal relationship between different stages: news and blog posts, social media mentions, citations in academic literature, and incorporation into policy and knowledge databases. We verified that for every variable, the distributions in the histograms are plausible. Because each observation relates to a distinct manuscript, our dataset contains 124,778 independent observations. Additionally, we looked for any anomalies or curvilinear relationships in the plotting of the dependent variable against each independent variable.

Note that although the relationship between the independent variables and policy citations is statistically significant, it is often small. As a result, all independent variables could be used simultaneously in the multiple linear regression model to explain policy citations. We then examined the relationships between the independent factors. As there are no high absolute correlations, multicollinearity problems are not taken into account in the regression analysis itself.

As mentioned above, the distributions of the variables are skewed and loaded with zeros (see Table 1). For this reason, we used robust ordinary least squares estimators. Each coefficient in Tables 4 and 5 indicates the average increase in policy citations associated with a one-unit increase in the predictors, all else being equal. However, the coefficients do not indicate the relative strength of the predictors. This is because the scales of the independent variables are different. The unpredictability of human behavior, which is often the subject of social science research, results in lower R-squared values. However, this does not detract from the value of the model, as the focus is on understanding the impact of certain variables rather than making precise predictions. In social science research, a small R-squared can be considered acceptable if most of the explanatory variables are statistically significant (Ozili, 2023). Furthermore, there is a non-zero correlation throughout the regression model because the p-value of the ANOVA is less than $10^{-4}$.

For each coefficient in Tables 4 and 5, the 2-tailed p-value can be found in the statistical significance column (Sig.). It is noteworthy that the majority of the coefficients in the models have high statistical significance, with p-values often less than $10^{-2}$. However, OA green in models 1 and 2, and OA gold and academic citations in model 2, do not have a significant impact. Both the year of publication of the article and the type of policy covered by the journal were found to be insignificant in both models, providing some robustness to the results obtained.

Factors that positively influence policy citations are news and blog mentions, social media mentions, Wikipedia mentions, academic citations (but only when modelling policy citations covered by Altmetric and with a very small effect), journal impact factor, funding and open access hybrid type. On the other hand, factors that negatively influence policy citations are paywalled (closed) access and open access gold type (but only when modelling policy citations covered by Altmetric). Finally, no evidence of an association with policy citations was found in either model for open access green type, year of publication and policy type.

In terms of effect size, we observed that for every 10 news items in the press or blogs, there is an increase of 1.8 policy citations in Overton and 0.4 in Altmetric. Conversely, every hundred social media interactions correspond to a rise in policy impact of 0.5 citations in Overton and 0.1 in Altmetric. Similarly, each ten mentions in Wikipedia pages is linked to an additional 1.8 policy citations in Overton and 0.5 in Altmetric. Furthermore, each extra point in a journal impact factor is associated with a 0.05 increase in policy citations in Overton and 0.02 in Altmetric. All these effects while keeping constant the rest of the variables (ceteris paribus).

At first sight, these effects may seem moderate, even small in some cases, compared to the possible effects of the same factors on scientific citations. However, two things can be said in this respect. The first is that the number of citations to political documents is lower than the number of citations to scientific documents (even in the most complete source, Overton). The second is that these effects may be cumulative. This means, for example, that the publication of an open access paper in a hybrid journal with an impact factor four points higher, which receives 10 references in the media, 100 interactions on social networks and 10 references on Wikipedia, is associated with a 4.3 increase in the number of policy citations (according to the Overton policy data source).

The standardized coefficients, shown in the last column of Tables 4 and 5, provide a more straightforward means of comparing the importance of different predictor variables in influencing the outcome variable, regardless of their original scales. This comparison is particularly useful when the predictors are measured using different scales or units, as is the case here. It is worth noting that the frequency of citations in academic documents far exceeds the frequency of mentions on Wikipedia pages, and the frequency of discussion of results on social media exceeds that of news and blogs, to name just two examples.

Based on these standardized coefficients, there are important differences between the predictors. Nevertheless, the results for the two models with different data sources for policy are similar in magnitude, with a few exceptions that we will discuss below. The most significant effects correspond to mentions in news and blogs (between 0.09 and 0.10, depending on the policy citation data source). In a second tier we find mentions in social media, with an effect of about a third of the later (ranging from 0.03 to 0.04, depending on the policy data source). A similar magnitude is observed for hybrid OA, although the estimation is inconsistent between the two models, with the effect size with Altmetric data being approximately double that with Overton data. In a third tier is the impact factor of the journal, with an effect quantifiable between 0.01 and 0.02, depending on the policy data source. In a fourth tier, considering the size of the effect, funding and mentions on Wikipedia are placed. In both cases, the effects are above 0.01 for both policy data sources. This tier could also include paywall access (closed OA) and open access to journals that publish all their content openly (gold OA), with absolute effect sizes above 0.01 in the case of the Altmetric data source. Finally, the smallest effect is observed in the case of academic citations, constituting a fifth tier in terms of effect size. In this case, focusing only on the significant case of the Altmetric data source, the effect size can be quantified below 0.01.

Therefore, among all of the predictors the effect of mentions in news and blogs stands out. This could be explained by the visibility associated with appearing in these media. It is worth noting that both science journalists and science communicators typically select the research they consider to be the most relevant in general and in the public policy field in particular. On the other hand, the surprising a priori fact that scientific citations have such a small effect compared to the other predictors highlights how far removed the knowledge referencing practices of policy researchers and policy makers are.

## 5. Discussion

The analysis of citations and social attention of articles in public policy journals (indexed in the Web of Science database) revealed distinctive implications for bibliographic referencing dynamics, particularly in the context of public policy. First, those involved in drafting policy documents exhibit referencing patterns that are markedly different from those of other stakeholders. Second, the referencing dynamics of researchers writing scientific papers differ from those of other actors, although they are comparable to those of Wikipedia contributors. Finally, referencing practices are similar between those involved in science communication journalism and blogging and general public conversations on social networks, but differ significantly from other stakeholder groups. These findings illustrate the diverse and complex ways in which different actors interact with and refer to scientific evidence, and they also highlight the importance of understanding these interactions in order to analyze how research influences public policy.

Combining different aspects that influence policy citations, we found that some components are essential for increasing the impact of scientific research on policy discourse. News and blog mentions are among the most important positive factors, as they act as a spark for initial public awareness and participation. Early exposure sparks the interest of a wider audience and lays the groundwork for later interactions in the political, societal and academic spheres (see Ortega, 2021). Social media mentions also emerge as a significant factor, reflecting the modern environment where digital platforms act as dynamic forums for the exchange of ideas. These exchanges take place in real time, speeding up the incorporation of research findings into wider public discourses and potentially influencing policy deliberations in the process (see Gong et al., 2023).

The value of Wikipedia mentions as a positive force is remarkable. The inclusion of research results in this openly accessible and collaboratively managed knowledge hub demonstrates a level of acceptance and recognition that extends beyond the scientific community (Mesgari et al., 2015). As a widely accessible platform, Wikipedia plays a crucial role in disseminating knowledge to a wide range of audiences, including professionals and policymakers. While it is true that academic citations have a positive correlation, it is crucial to consider the results for Altmetric policy data, which show a small effect size for policy citations. This suggests that while citations contribute to the scholarly impact of a paper, their role in shaping policy discussions may be more nuanced and indirect (see Bornmann et al., 2016).

The journal impact factor is a relevant indicator of the importance of the publication venue in determining the influence and impact of research. Journals with a higher impact factor are not only better recognized by policymakers, but also receive more attention from the academic community. In addition, funding plays a crucial role in demonstrating institutional and financial support (Dorta-González & Dorta-González, 2023b). The fact that it is positively associated with policy citations suggests that research backed by strong funding mechanisms is more likely to attract the interest of policymakers.

It has been observed that certain accessibility factors can have a significant impact on policy citation rates. A prime example is closed access, where publication in paywalled journals can act as a barrier to policymakers. In addition, research made available through open access channels, particularly hybrid models, has a greater potential to bridge the gap between policy considerations and academic evidence. These results confirm others previously obtained in the case of academic citations (see Dorta-González & Dorta-González, 2023a,b). Interestingly, no clear relationship was found between policy citations and accessibility through institutional or thematic open repositories. This lack of association highlights the need for further research into the multiple effects of different access models on policy engagement. On the other hand, both the year of publication of the paper and the type of policy were found to be non-significant in both policy data sources, providing some robustness to the results obtained.

In conclusion, policy citations are shaped by a complex interplay between scientific research and its impact on society through policy discourse. Our findings enhance our understanding of these dynamics and provide valuable insights for researchers, policymakers, and stakeholders seeking to optimize the practical application of scientific knowledge. These insights deepen our understanding of policy impact. They also have practical significance for researchers, policymakers, and communicators who aim to enhance the breadth and efficacy of their contributions in shaping public policy agendas.

Acknowledging limitations in our approach is crucial, especially when using news articles and blog posts as sources of data. Traditional press outlets tend to prioritize sensational aspects of scientific research, which can result in an imbalance in coverage across different research areas. As a result, some fields may receive less attention from mainstream media sources than others. Scientists often disseminate information through specialized blogs. However, it is important to recognize that these platforms may not reach as broad of an audience as traditional news outlets. Therefore, we opted to incorporate both news articles and blog posts in our analysis. This is because they share the objective of reaching a wider audience beyond the realm of specialized academia. However, it is important to acknowledge that this decision may introduce biases in our dataset. These biases could potentially influence the interpretation and generalizability of our results. Future research efforts should aim to address these limitations and explore alternative methodologies to ensure a more comprehensive and balanced representation of scientific communication in the media landscape.

We used two large databases to measure how scientific research influences policymaking. Overton identified many more citations than Altmetric, which could be due to less substantial mentions in Overton. By combining these sources and using different analysis techniques, the study aimed to provide a more robust picture of how research and policy interact around the world.

As a final consideration, the striking finding that scientific citations have a disproportionately low impact compared to other predictors highlights the profound disconnect between the referencing practices of policy researchers and the decision-making processes of policy makers. This finding underscores a broader issue within the field of policy formulation and implementation, suggesting that while scientific research serves as a foundational pillar of evidence-based policymaking, its direct impact on shaping policy outcomes may be limited.

This observation prompts critical reflection on the dynamics between science and policy, and points to potential gaps in the communication, interpretation or application of scientific evidence in policy contexts. It raises questions about the effectiveness of current knowledge dissemination strategies and the extent to which policy makers engage with and integrate academic research into their decision-making frameworks. Moreover, this gap underscores the need for more nuanced approaches to bridging the gap between research and policy, emphasizing not only the importance of generating robust scientific evidence, but also the imperative of fostering meaningful dialogue and collaboration between researchers and policymakers. Addressing this gap may require interdisciplinary efforts aimed at improving the accessibility, relevance and applicability of research findings to the complex and dynamic landscape of policy formulation and implementation.

While numerous methods exist to assess the non-academic impact of research (e.g., surveys), we deliberately focused on citations within policy documents. This approach offers a strategic lens into the policy sphere, enabling us to pinpoint research articles demonstrably influencing policy development. This focus on policy citations presents several advantages. Firstly, it provides a quantifiable measure of impact, facilitating robust statistical analysis and comparisons across diverse research fields and policy areas. This allows us to identify trends and patterns in how policymakers utilize research evidence.

However, it's crucial to acknowledge potential limitations inherent to this approach. Policy documents might not capture all instances where research informs policy. Additionally, relying solely on citations within these documents introduces potential biases. For example, some policy areas might exhibit a stronger culture of citing research compared to others. To address these limitations and gain a more nuanced understanding of research-policy interactions, future studies could employ complementary methods. Interviews with policymakers or analyses of policy deliberations could provide valuable insights into the intricate dynamics of knowledge transfer between academia and the policymaking world.

**Acknowledgement**

The authors would like to thank to the Editor and three anonymous referees for valuable comments and suggestions which improved the presentation of the paper.

Table 1. Statistical data derived from a sample size of 124,778 observations

|  | Min. | 1st Qu. | Median | Mean | 3rd Qu. | Max. |
|---|---|---|---|---|---|---|
| 1 News and Blog mentions | 0 | 0 | 0 | 0.7 | 0 | 396 |
| 2 Social Media mentions | 0 | 0 | 1 | 7.7 | 5 | 11,991 |
| 3 Wikipedia mentions | 0 | 0 | 0 | 0.1 | 0 | 55 |
| 4 Academic citations | 0 | 3 | 12 | 31.9 | 32 | 11,895 |
| 5 Policy citations in Altmetric | 0 | 0 | 0 | 0.6 | 0 | 110 |
| 6 Policy citations in Overton | 0 | 0 | 0 | 2.3 | 2 | 490 |

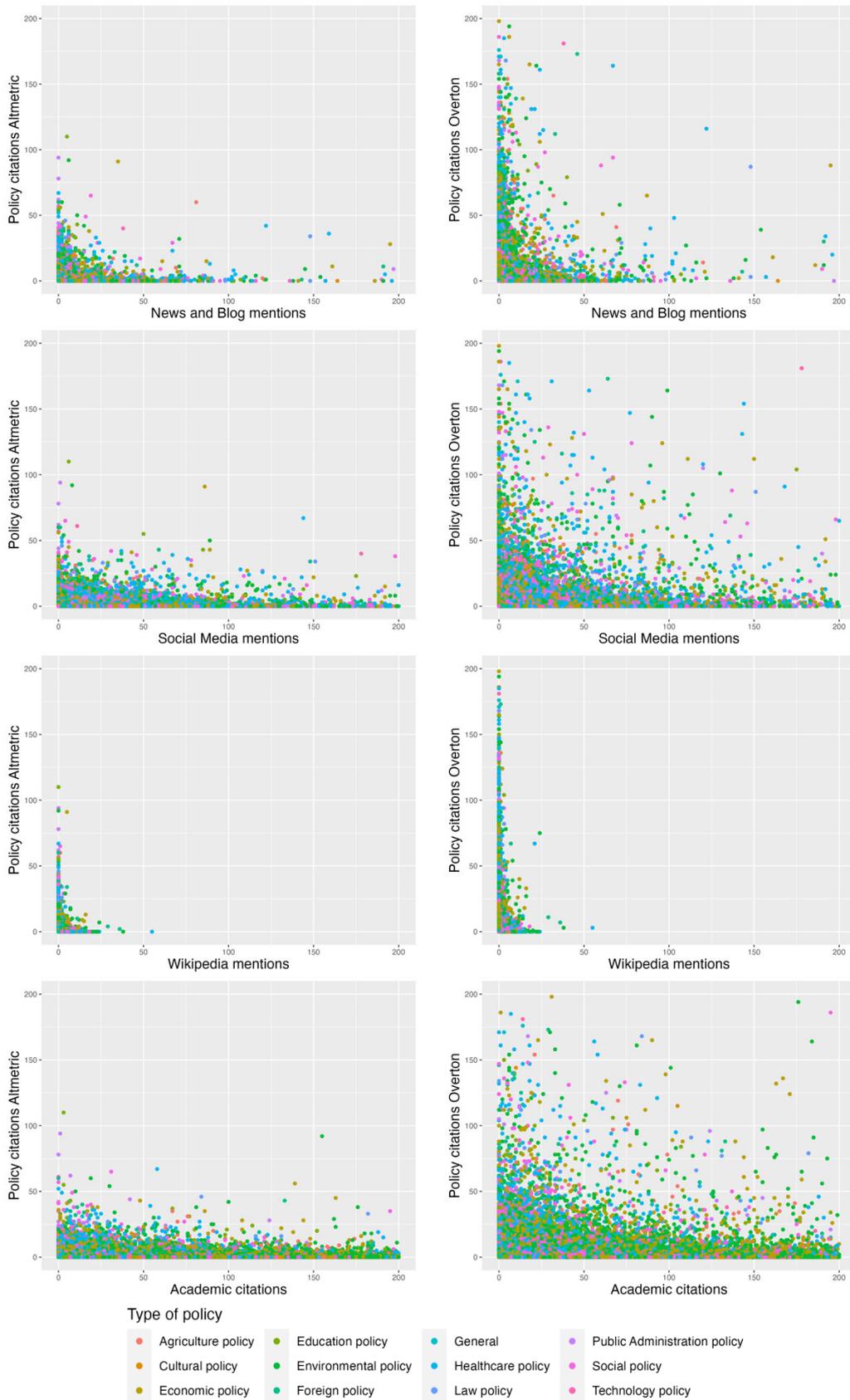

Figure 1. Scatterplot between policy citations in Altmetric (left), Overton (right) and the rest of the variables, by type of policy

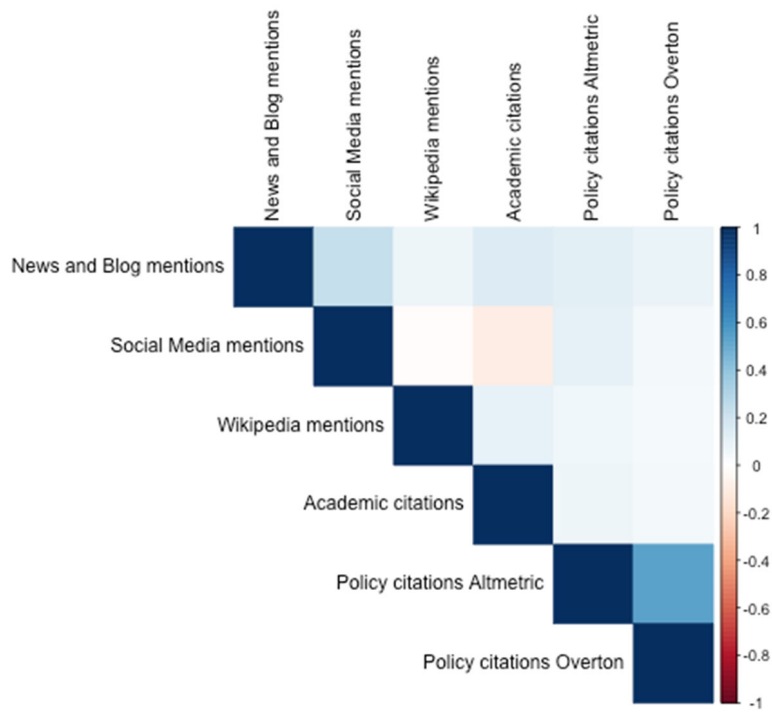

Figure 2. Spearman correlations

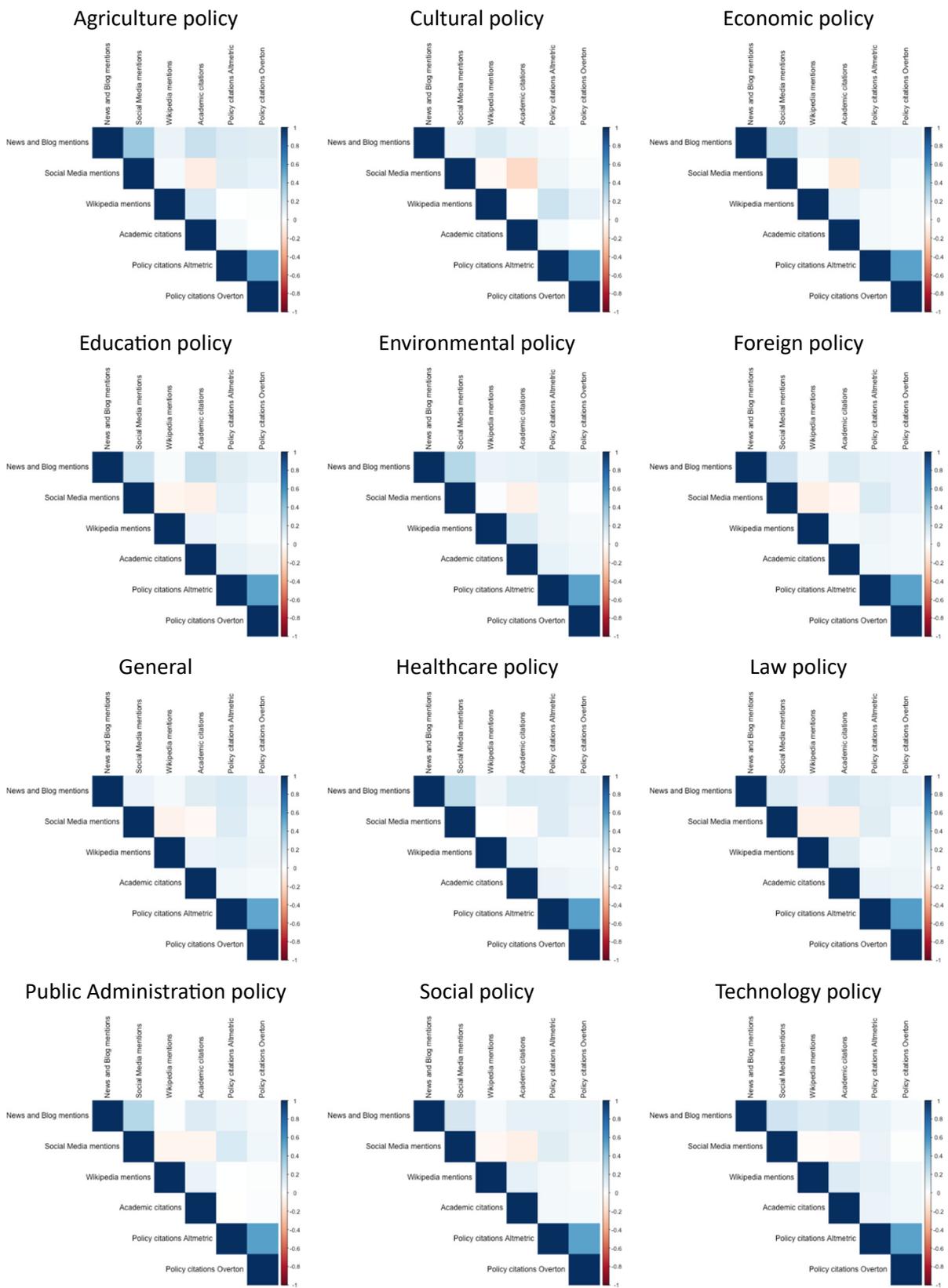

Figure 3. Spearman correlations by type of policy

Table 2. PCA summary. Importance of components

|  | PCA1 | PCA2 | PCA3 | PCA4 | PCA5 | PCA6 |
|---|---|---|---|---|---|---|
| Standard deviation | 1.2980 | 1.1334 | 1.0195 | 0.9109 | 0.8717 | 0.6329 |
| Proportion of Variance | 0.2810 | 0.2141 | 0.1732 | 0.1383 | 0.1266 | 0.0668 |
| Cumulative Proportion | 0.2810 | 0.4951 | 0.6683 | 0.8066 | 0.9333 | 1 |

Table 3. Loading matrix of the first three principal components

|  | PCA1 | PCA2 | PCA3 |
|---|---|---|---|
| 1 News and Blog mentions | -0.320 | -0.458 | 0.340 |
| 2 Social Media mentions | -0.230 | -0.386 | 0.636 |
| 3 Wikipedia mentions | -0.163 | -0.497 | -0.410 |
| 4 Academic citations | -0.136 | -0.462 | -0.544 |
| 5 Policy citations in Altmetric | -0.630 | 0.304 | -0.101 |
| 6 Policy citations in Overton | -0.634 | 0.296 | -0.081 |

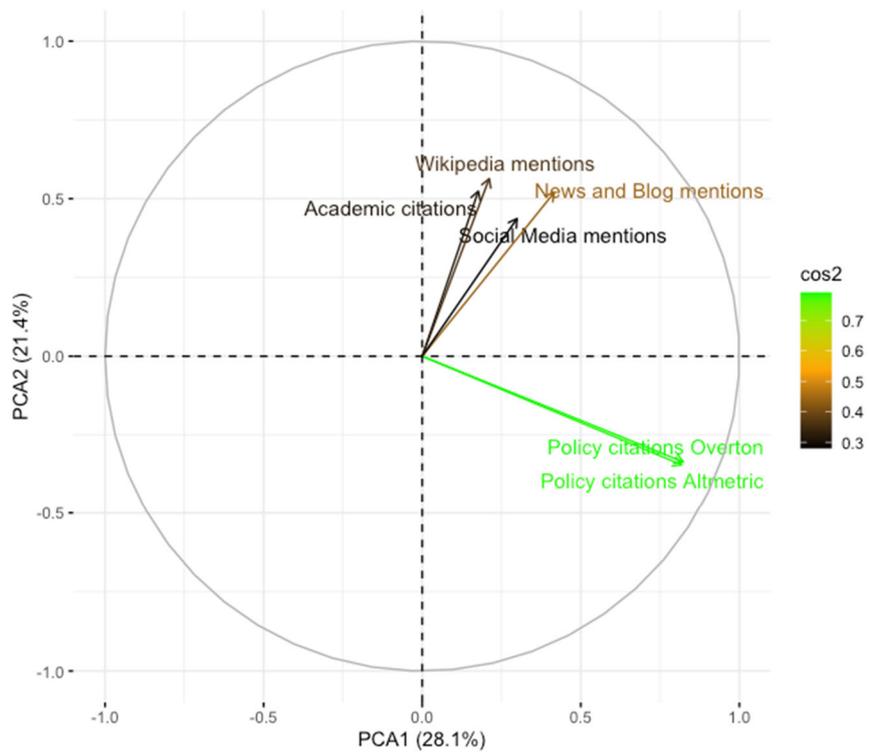

Figure 4. Biplot showing the contribution of the variables to the first two principal components together with the square cosine (cos2) values

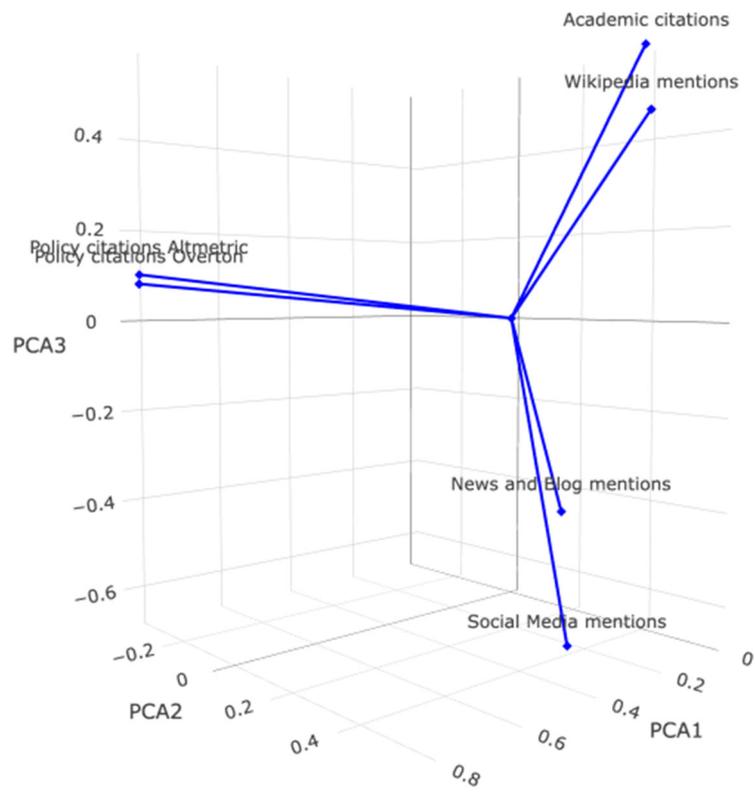

Figure 5. Three-dimensional plot showing the contribution of the variables to the first three principal components

Table 4. Effect of the explanatory variables on policy citations in Altmetric (OLS robust Model 1)

|  | Coef. | Std. Err. | t | Sig. (P>t) | [95% conf. | interval] | Std. Coef. |
|---|---|---|---|---|---|---|---|
| News and Blog mentions | 0.0374 | 0.0054 | 6.9900 | 0.0000 | 0.0269 | 0.0479 | 0.0887 |
| Social Media mentions | 0.0009 | 0.0004 | 2.1500 | 0.0320 | 0.0001 | 0.0017 | 0.0295 |
| Wikipedia mentions | 0.0462 | 0.0149 | 3.1000 | 0.0020 | 0.0170 | 0.0754 | 0.0145 |
| Academic citations | 0.0002 | 0.0001 | 2.0400 | 0.0410 | 0.0000 | 0.0004 | 0.0088 |
| Journal Impact Factor | 0.0195 | 0.0028 | 6.9700 | 0.0000 | 0.0140 | 0.0250 | 0.0240 |
| Funding | 0.0638 | 0.0142 | 4.5000 | 0.0000 | 0.0360 | 0.0916 | 0.0143 |
| OA type (respect to OA bronze) | | | | | | | |
| OA closed | -0.0578 | 0.0257 | -2.2500 | 0.0240 | -0.1081 | -0.0075 | -0.0138 |
| OA gold | -0.0853 | 0.0322 | -2.6400 | 0.0080 | -0.1485 | -0.0221 | -0.0111 |
| OA green | 0.0184 | 0.0295 | 0.6300 | 0.5320 | -0.0393 | 0.0762 | 0.0032 |
| OA hybrid | 0.2182 | 0.0364 | 5.9900 | 0.0000 | 0.1468 | 0.2895 | 0.0313 |
| Type of Policy (respect to Agriculture policy) | | | | | | | |
| Cultural policy | -0.1838 | 0.0756 | -2.4300 | 0.0150 | -0.3319 | -0.0357 | -0.0078 |
| Economic policy | -0.1770 | 0.0545 | -3.2500 | 0.0010 | -0.2838 | -0.0702 | -0.0340 |
| Education policy | -0.0558 | 0.0631 | -0.8800 | 0.3760 | -0.1794 | 0.0678 | -0.0063 |
| Environmental policy | -0.2023 | 0.0540 | -3.7500 | 0.0000 | -0.3082 | -0.0964 | -0.0418 |
| Foreign policy | -0.0841 | 0.0586 | -1.4400 | 0.1510 | -0.1988 | 0.0307 | -0.0104 |
| General | -0.0918 | 0.0659 | -1.3900 | 0.1640 | -0.2209 | 0.0373 | -0.0067 |
| Healthcare policy | -0.0829 | 0.0555 | -1.4900 | 0.1350 | -0.1917 | 0.0258 | -0.0162 |
| Law policy | -0.1273 | 0.0615 | -2.0700 | 0.0380 | -0.2479 | -0.0068 | -0.0114 |
| Public Administration policy | -0.0079 | 0.0687 | -0.1200 | 0.9080 | -0.1426 | 0.1267 | -0.0007 |
| Social policy | -0.0891 | 0.0568 | -1.5700 | 0.1170 | -0.2005 | 0.0223 | -0.0138 |
| Technology policy | -0.2693 | 0.0609 | -4.4200 | 0.0000 | -0.3887 | -0.1500 | -0.0178 |
| Cons | 0.5943 | 0.1911 | 3.1100 | 0.0020 | 0.2198 | 0.9688 | . |
| Number of obs. | 122,837 | | | | | | |
| F(78, 122758) | 25.08 | | | | | | |
| Prob > F | $10^{-4}$ | | | | | | |

Table 5. Effect of the explanatory variables on policy citations in Overton (OLS robust Model 2)

|  | Coef. | Std. Err. | t | Sig. (P>t) | [95% conf. | interval] | Std. Coef. |
|---|---|---|---|---|---|---|---|
| News and Blog mentions | 0.1817 | 0.0257 | 7.0700 | 0.0000 | 0.1313 | 0.2320 | 0.1027 |
| Social Media mentions | 0.0047 | 0.0022 | 2.1300 | 0.0330 | 0.0004 | 0.0089 | 0.0380 |
| Wikipedia mentions | 0.1778 | 0.0669 | 2.6600 | 0.0080 | 0.0467 | 0.3089 | 0.0133 |
| Academic citations | 0.0000 | 0.0003 | 0.1600 | 0.8710 | -0.0005 | 0.0006 | 0.0005 |
| Journal Impact Factor | 0.0516 | 0.0116 | 4.4600 | 0.0000 | 0.0289 | 0.0742 | 0.0151 |
| Funding | 0.2437 | 0.0599 | 4.0700 | 0.0000 | 0.1264 | 0.3611 | 0.0130 |
| OA type (respect to OA bronze) | | | | | | | |
| OA closed | -0.2273 | 0.1108 | -2.0500 | 0.0400 | -0.4445 | -0.0102 | -0.0129 |
| OA gold | -0.2092 | 0.1431 | -1.4600 | 0.1440 | -0.4897 | 0.0714 | -0.0065 |
| OA green | 0.0115 | 0.1274 | 0.0900 | 0.9280 | -0.2381 | 0.2612 | 0.0005 |
| OA hybrid | 0.4766 | 0.1519 | 3.1400 | 0.0020 | 0.1788 | 0.7744 | 0.0163 |
| Type of Policy (respect to Agriculture policy) | | | | | | | |
| Cultural policy | -0.0359 | 0.3368 | -0.1100 | 0.9150 | -0.6962 | 0.6243 | -0.0004 |
| Economic policy | -0.2790 | 0.2084 | -1.3400 | 0.1810 | -0.6874 | 0.1295 | -0.0128 |
| Education policy | -0.0914 | 0.2281 | -0.4000 | 0.6890 | -0.5385 | 0.3556 | -0.0025 |
| Environmental policy | -0.3659 | 0.2033 | -1.8000 | 0.0720 | -0.7644 | 0.0326 | -0.0181 |
| Foreign policy | -0.1777 | 0.2222 | -0.8000 | 0.4240 | -0.6131 | 0.2577 | -0.0052 |
| General | 0.0395 | 0.2793 | 0.1400 | 0.8880 | -0.5079 | 0.5869 | 0.0007 |
| Healthcare policy | -0.0994 | 0.2118 | -0.4700 | 0.6390 | -0.5145 | 0.3157 | -0.0046 |
| Law policy | -0.2574 | 0.2314 | -1.1100 | 0.2660 | -0.7109 | 0.1961 | -0.0055 |
| Public Administration policy | 0.1832 | 0.2862 | 0.6400 | 0.5220 | -0.3778 | 0.7441 | 0.0041 |
| Social policy | -0.1427 | 0.2151 | -0.6600 | 0.5070 | -0.5643 | 0.2788 | -0.0053 |
| Technology policy | -0.5768 | 0.2326 | -2.4800 | 0.0130 | -1.0326 | -0.1209 | -0.0091 |
| Cons | 1.7035 | 0.5948 | 2.8600 | 0.0040 | 0.5376 | 2.8694 | . |
| Number of obs. | 122,837 | | | | | | |
| F(78, 122758) | 26.96 | | | | | | |
| Prob > F | $10^{-4}$ | | | | | | |